\documentclass[aip,jcp,reprint,10pt]{revtex4-1}

\usepackage{dcolumn}       
\usepackage{bm}            
 \usepackage{times}

\usepackage{amsmath}       
\usepackage{graphicx}      
\usepackage{natbib}		   
\usepackage{hyperref}      
\usepackage{amssymb}       
\usepackage{latexsym}      
\usepackage{array}         
\usepackage{epstopdf}     

\newcommand{\bs}[1] {  \boldsymbol{#1}           }
\newcommand{\ff}[1] {  \mbox{\footnotesize{#1}}  }
\newcommand{\Ang}   {  \mbox{\normalfont\AA}     }

\begin{document}

\title{ Using a monomer potential energy surface to perform approximate path integral molecular dynamics simulation of ab initio water at near-zero added cost}

\author{Daniel C.\ Elton}
\affiliation{Department of Physics and Astronomy, Stony Brook University, Stony Brook, New York 11794-3800, USA}
\affiliation{Institute for Advanced Computational Sciences, Stony Brook University, Stony Brook, New York 11794-3800, USA}
\email{delton@umd.edu}
\author{Michelle Fritz}
\affiliation{Universidad Autonoma de Madrid, 28049 Madrid, Spain.}
\author{M.-V. Fern\'{a}ndez-Serra}
\email{maria.fernandez-serra@stonybrook.edu}
\affiliation{Department of Physics and Astronomy, Stony Brook University, Stony Brook, New York 11794-3800, USA}
\affiliation{Institute for Advanced Computational Sciences, Stony Brook University, Stony Brook, New York 11794-3800, USA}

\date{\today}
\begin{abstract}
It is now established that nuclear quantum motion plays an important role in determining water's hydrogen bonding, structure, and dynamics. Such effects are important to include in density functional theory (DFT) based molecular dynamics simulation of water. The standard way of treating nuclear quantum effects, path integral molecular dynamics (PIMD), multiplies the number of energy/force calculations by the number of beads required. In this work we introduce a method whereby PIMD can be incorporated into a DFT simulation with little extra cost and little loss in accuracy. The method is based on the many body expansion of the energy and has the benefit of including a monomer level correction to the DFT energy. Our method calculates intramolecular forces using the highly accurate monomer potential energy surface developed by Partridge-Schwenke, which is cheap to evaluate. Intermolecular forces and energies are calculated with DFT only once per timestep using the centroid positions. We show how our method may be used in conjunction with a multiple time step algorithm for an additional speedup and how it relates to ring polymer contraction and other schemes that have been introduced recently to speed up PIMD simulations. We show that our method, which we call ``monomer PIMD'', correctly captures changes in the structure of water found in a full PIMD simulation but at much lower computational cost.
\end{abstract}

\maketitle

There is great interest in being able to accurately simulate liquid water at the quantum mechanical level.\cite{Gaiduk2015:2902,Guardia2015:8926,Corsetti2013:194502,Gillan2016:130901} The most widely used methodology for this is density functional theory. However, many density functionals fail to accurately reproduce all of the key properties of water such as its density, compressibility, and diffusion constant. Moreover, different density functionals fail in different ways. For instance, PBE creates a overstructured liquid, while many van der Waals (vdW) functionals create an understructured liquid.\cite{Wang2011:024516, Mogelgoj2011:14149} There are nonetheless new meta-GGA functionals such as SCAN\cite{Chen2017:SCANPNAS} or empirically optimized hybrid functionals such as  B97M-rV\cite{Pestana2018:TheQuestJPhysChemLett} which are producing promising results for liquid water.

Most ab initio techniques are based on the Born-Oppenheimer approximation and the assumption that nuclear dynamics can be treated classically. However, over the past two decades a wide range of studies have demonstrated that this is not a good assumption for water because the OH stretching mode of water is very quantum mechanical (zero point temperature $T_z = \hbar\omega/2k_b = 2600$ K), and hydrogen nuclei are delocalized, leading to a large number of non-negligible nuclear quantum effects (NQEs) -- for a recent review, see Ceriotti, et al.\cite{Ceriotti2016review}


In the primary isotope effect, the OH distance is observed to be longer than the OD distance. In the secondary isotope effect, also called the Ubbel\"{o}hde effect, the H-bond donor-acceptor (oxygen-oxygen) distance $R$ changes upon isotopic substitution. The magnitude and direction of the change depends on the strength of the hydrogen bond, due to competing quantum effects.\cite{Kim2017:075502, McKenzie2014:174508,Habershon2009:024501,Li2011:6369,Markland2012:7988,Romanelli2013:3251} In particular, the zero-point motion of hydrogen in the out-of-plane direction (a type of librational motion) acts to increase $R$ while the zero point motion of the stretching mode acts to decrease $R$.\cite{McKenzie2014:174508}

In materials with strong H-bonds, NQEs decrease the donor-acceptor distance (positive Ubbel\"{o}hde effect), while in materials with weaker H-bonds the opposite effect occurs (negative Ubbel\"{o}hde effect). The crossover from positive to negative Ubbel\"{o}hde effect has been estimated to be around $R = 2.6-2.7\ \Ang$.\cite{Li2011:6369, McKenzie2014:174508} Both water and ice have H-bonds that lie near this crossover point,\cite{Gillan2016:130901} and therefore the magnitude and direction of the secondary isotope effect in simulations of water and ice is particularly sensitive to the details of the water geometry. The secondary isotope effect in ice is known to be positive (NQEs decrease $R$), leading to the anomalous isotope effects discovered by Pamuk et al.\cite{Pamuk2012:193003,Pamuk2015:134105} The anomalous isotope effect occurs in several phases of ice and persists even in room temperature water.\cite{Pamuk2018insights}


The ``gold standard'' technique for treating NQEs is path integral molecular dynamics (PIMD).\cite{Chandler1981:4078}
The sensitivity of competing quantum effects to the water geometry and and degree of anharmonicity in the OH potential leads to a broad spectrum of sometimes conflicting results obtained from PIMD simulations of water with different forcefield models and DFT functionals.\cite{Chen2003:215503,Li2011:6369,McKenzie2014:174508} As an example, the change in the dipole moment of H$_2$O when NQEs are included may be either positive or negative depending on the functional or forcefield being employed.\cite{Habershon2009:024501,EltonThesis} Because of the high cost of incorporating NQEs with PIMD, the testing of DFT functionals is often done with D$_2$O, where NQEs are much smaller due to the higher mass of deuterium and can therefore be ignored. This may be reasonable for testing density functionals, but the structure and dynamics of D$_2$O are different than H$_2$O due to NQEs. In the past, some people have introduced ``effective NQEs" by raising the temperature of their DFT simulation. This can be justified theoretically for weakly interacting systems such as gases or van der Waals bonding materials,\cite{landau1969statistical} but the same justification does not apply to hydrogen bonded materials. Increasing the temperature can be useful for compensating for the overstructuring of GGA functionals, but should not considered as an effective treatment of NQEs. A better option for approximately simulating NQEs is to use colored noise thermostats tuned to quantum zero point temperatures of different modes in liquid water.\cite{Ganeshan2013:134207}

We note that classical forcefield models are not a rigorous way of studying NQEs because they are parametrized to experimental data, leading to a double counting of NQEs when used with PIMD simulation. Additionally, harmonic models do not allow for a change in the average OH distance from NQEs, and thus cannot capture primary or secondary isotope effects. Even worse, we have found that PIMD simulation with the harmonic model SPC-f\cite{Toukan1985:2643} (and to a lesser extent q-SPC/Fw) shows an unphysical \textit{decrease} in $r_{OH}$,\cite{EltonThesis} which must be due to the ``curvature problem" intrinsic to PIMD simulation. In the curvature problem, beads curve around a spherical shell of near constant $r_{OH}$, causing the centroid to lie in the interior, leading to a shorter $r_{OH}$.\cite{Witt:194510,Paesani:014105,EltonThesis}  While classical forcefields have been reparametrized specifically for use with PIMD,\cite{Fritsch2014:816,john2015quantum} and also parametrized from Born Oppenheimer ab initio simulations,\cite{Spura2015:808,Paesani2006:184507} the most rigorous and computationally attainable way of
studying NQEs in liquid water is by means of DFT-based PIMD simulations.

-------------------------------------------------
\section{Path integral molecular dynamics methods}

PIMD maps the partition function for the quantum mechanical system onto the partition function of a classical system with the following Hamiltonian:
\begin{equation}
    \begin{aligned}
	H = &  \sum\limits_{i=1}^N \sum\limits_{j=1}^{N_b} \left( \frac{(\bs{p}_i^j)^2}{2 m_i'} + \frac{m_i\omega_n^2}{2}(\bs{q}_i^j - \bs{q}_i^{j+1})^2 \right) \\
	&+ \sum\limits_{j=1}^{N_b }V(\bs{q}_1^j, \cdots, \bs{q}_N^j )
	\end{aligned}
\end{equation}
Here $q_i^k$ are (x,y,z) vectors containing the bead coordinates and $i = 1...N$ is the atomic index and $k = 1...N_b$ is the bead index. We have put a prime on $m_i'$ to indicate that these masses (called fictitious masses) may be different than the physical masses $m_i$. A full derivation and description of the PIMD method can be found elsewhere.\cite{EltonThesis} Craig and Manolopoulos argue that simply setting $m_i' = m_i$ for all $i$ does the best job of reproducing the actual quantum dynamics, and call this methodology ``Ring Polymer Molecular Dyanmics" (RPMD). However, when RPMD is used the spectra are contaminated by spurious peaks caused by the normal mode frequencies which span the entire spectrum from $0$ to $2\omega_n$ where $\omega_n = k_B T N_b / \hbar$.\cite{Rossi2014:234116,Habershon2008:074501}

In this section we discuss different options for rescaling the fictitious masses as $m'_j = \sigma_j m_j$, where $\sigma_j$ is a ``mass rescaling factor". The rescaling is typically done in normal mode coordinates, so $j$ indexes the normal modes of the ring polymer. The mass rescaling factor rescales the bead normal mode frequencies as $\Omega_k' = \Omega_k/\sqrt{\sigma_k}$. Ignoring thermostating choices, the major different PIMD implementations that have been introduced are distinguished solely by their choice of mass rescaling factor.\cite{Witt:194510,EltonThesis}

\begin{figure}\centering
    \includegraphics[width=8.5cm]{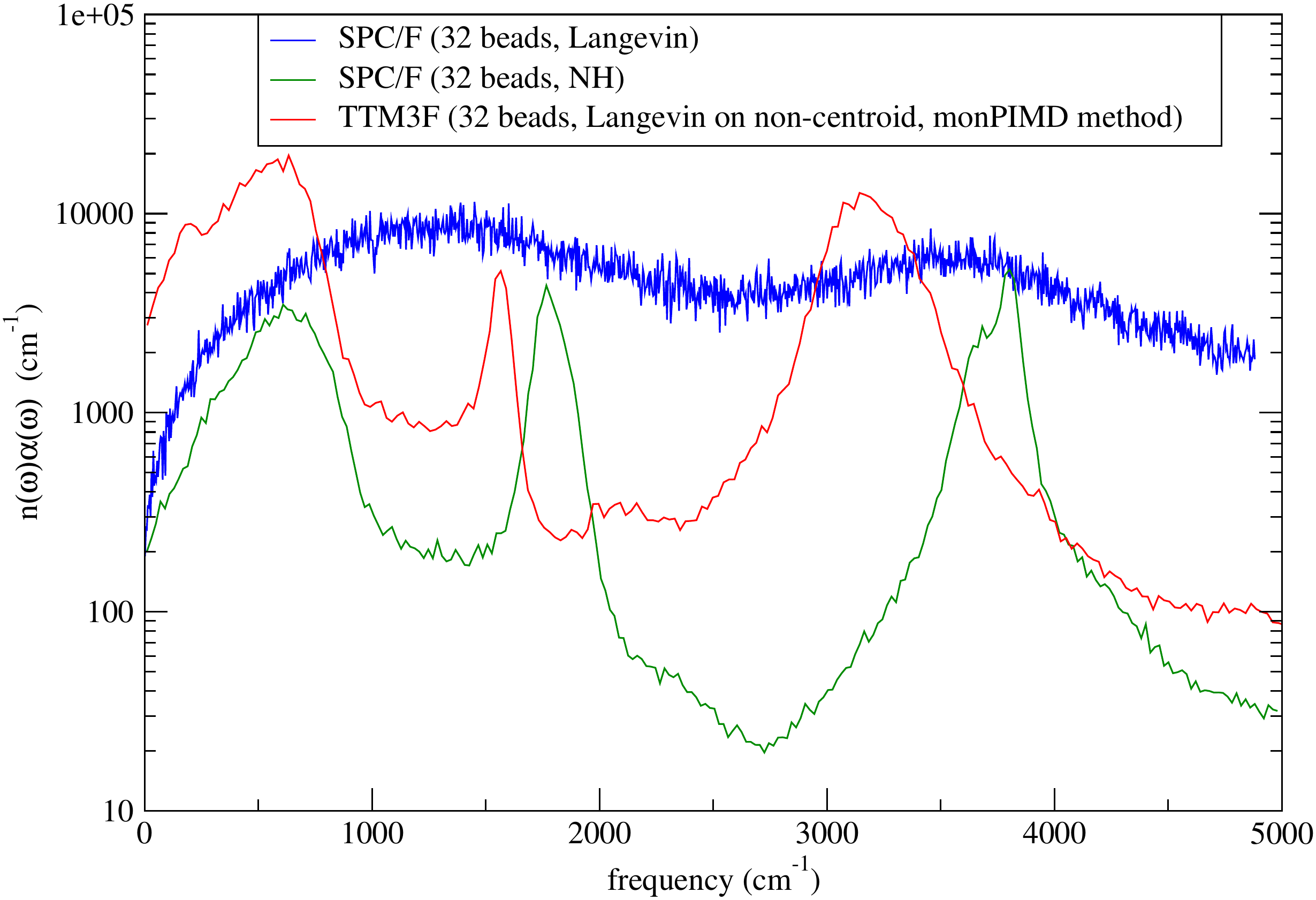}
    \caption{Comparison of IR spectra using Langevin (PILE) and Nos\'{e}-Hoover thermostating. The IR spectra from PIMD simulation are shown for SPC/F with Langevin thermostating on all the modes, which washes out the dynamics. We do not thermostat the centroid mode with PILE, which preserves the dynamics, as shown for TTM3F and the monomer PIMD method. }
    \label{thermostatcompare}
\end{figure}
In their original paper on PIMD,\cite{Parrinello:860} Parrinello et al.\ choose to bring all of the non-centroid frequencies to the value of $\omega_n$. A better approach is to scale the frequencies of the normal modes to above the highest frequency of interest in the system, thus avoiding the problem of normal mode contamination.\cite{Habershon2008:074501,Hone:154103}
In effect, centroid molecular dynamics (CMD) rescales the normal modes to a very high frequency.\cite{Jang1999:2357} The disadvantage of this is that it requires using a very short timestep, even when an exact propagator is used to evolve the normal mode coordinates. The PIMD simulation methodology we use is called ``partially adiabatic centroid molecular dynamics", denoted PA-CMD, because we choose an intermediate rescaling.\cite{Hone:154103,EltonThesis} In most of our work we scale all normal modes to 10,000 cm$^{-1}$, well above the overtones found at 5260 cm$^{-1}$ and 6800 cm$^{-1}$.

The other ingredient to PIMD is to attach thermostats to each degree of freedom to overcome the ergodicity problems first pointed out by Hall and Berne (1984).\cite{Hall:3641} We use Nos\'{e}-Hoover chain thermostats, with a chain length of 2. Alternatively, our code allows for Langevin thermostats to be used. The thermostating is done in normal-mode space, with the thermostats optimally tuned to each normal mode as they are in the PILE thermostat scheme of Ceriotti et al.\cite{Ceriotti2010:124104} Importantly, the centroid mode is not thermostated, since doing so washes out the dynamics (as shown in fig.\ \ref{thermostatcompare}).

\begin{figure}\centering
    \includegraphics[width=8cm]{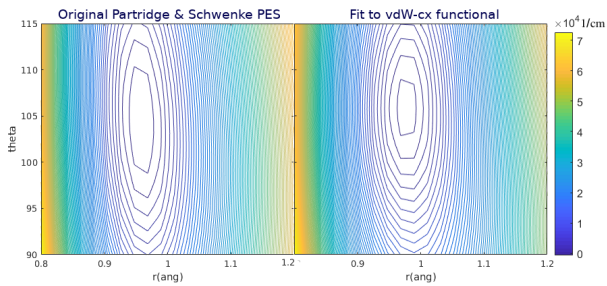}
    \caption{The monomer potential energy surface of Partridge and Schwenke (left) and vdW-cx (right).}
    \label{PESfit}
\end{figure}

\subsection{The many body expansion}
Our method is based on the many body expansion, which gives an exact decomposition of the potential energy into 1-body, 2-body, 3-body, and higher order terms:
\begin{equation}
	\begin{aligned}
    V(\lbrace\bs{R}_I\rbrace) =& \sum\limits_{I=1}^{N_{\ff{mol}}} V_1(\bs{R}_I) + \sum\limits_{I<J}^{N_{\ff{mol}}} V_2(\bs{R}_I,\bs{R}_J) \\
    &+ \sum\limits_{I<J<K}^{N_{\ff{mol}}} V_3(\bs{R}_I,\bs{R}_J,\bs{R}_K) + \cdots
    \end{aligned}
\end{equation}
Here $R_I$ refers to the set of nuclear coordinates of molecule $I$, and $N_{\ff{mol}}$ is the number of molecules. In our method, we first subtract off the DFT monomer energies using a monomer potential energy surface (described below) fitted to the DFT functional being used. By subtracting off this term, this allows us to calculate the intramolecular energy using PIMD with the Partridge-Schwenke monomer potential energy surface (PES),\cite{Partridge1997:4618} which is a highly accurate surface derived from CCSD calculations. This can thought of as a monomer correction to the DFT potential:
\begin{equation}\label{monomersubtr}
    V'(\lbrace\bs{R}_I\rbrace) = V_{\ff{DFT}}(\lbrace\bs{R}_I\rbrace) - \sum\limits_{I=1}^{N_{\ff{mol}}} V_{\ff{1DFT}}(\bs{R}_I) + \sum\limits_{I=1}^{N_{\ff{mol}}} V_{\ff{1PS}}^{\ff{(PIMD)}}(\bs{R}_I)
\end{equation}

Therefore, the intramolecular energies and forces are calculated with PIMD, while the intermolecular forces and energies are calculated using standard techniques. The intermolecular forces on the beads are all set equal to the intermolecular forces computed from the bead centroids. Thus, in each timestep we only have to do one DFT calculation, using the centroid coordinates.

\begin{figure}\centering
 \centering
 \includegraphics[width=8.5cm]{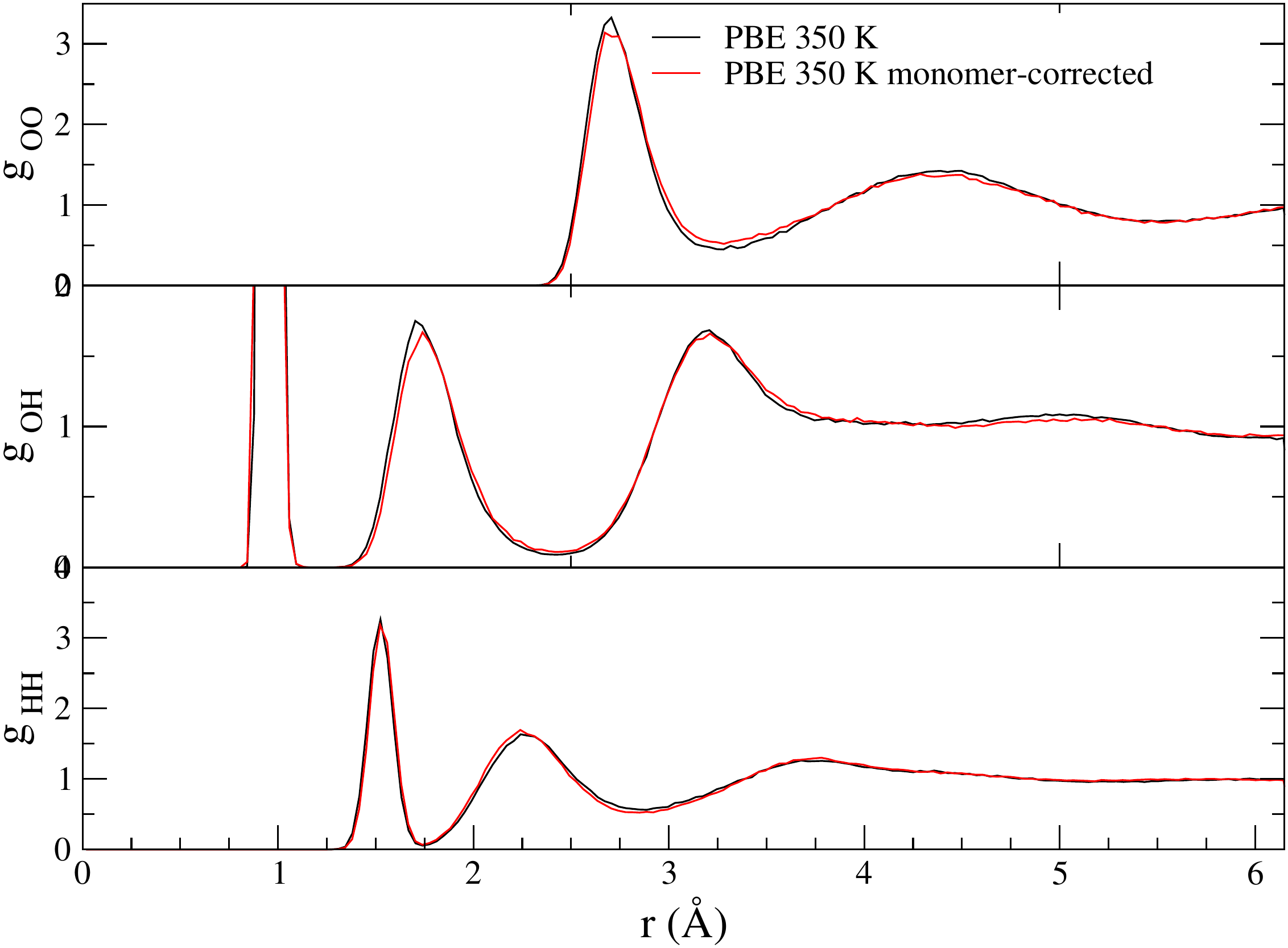}
  \caption{Comparison of RDFs for conventional PBE and monomer corrected PBE. The simulations had 64 molecules and lengths of 35 and 27 ps, respectively.}
  \label{PBEmoncorrRDFs}
\end{figure}

In addition to allowing for more efficient calculation of NQEs, our method has the added advantage of including a monomer correction to the DFT energy.\cite{Fritz2016:224101} It has previously been shown that a large contribution to DFT error is in the monomer term.\cite{Bartok2013:054104} A comparison of radial distribution functions (RDFs) for conventional PBE and monomer-corrected PBE with 64 molecules is shown in figure \ref{PBEmoncorrRDFs}. It is worth noting that in place of a monomer correction, the PES fit to the functional being used may be used instead, as may be desired for doing a rigorous comparison of different functionals with our method.

\subsection{Monomer potential energy surface}
\begin{figure}\centering
 \includegraphics[width=8.5cm]{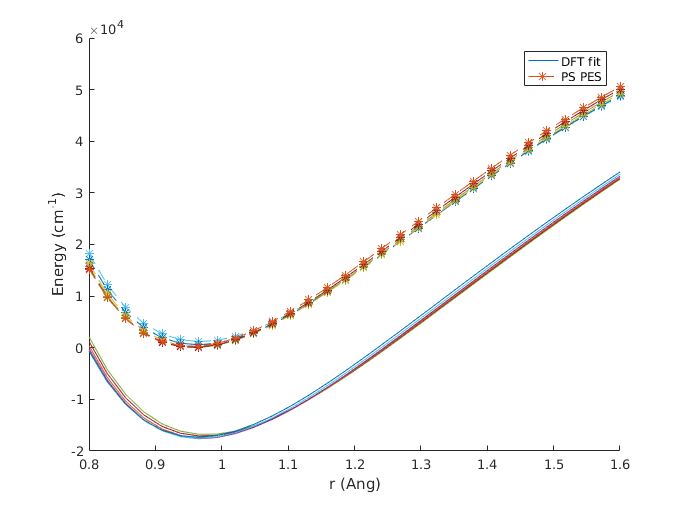}\\
  \caption{Energy vs $r_{\ff{OH}}$ for the case where $r_{\ff{OH1}} = r_{\ff{OH2}}$. Different HOH angles are shown in different colors. The Partridge and Schwenke energy surface is compared with a custom fit to PBE.}
   \label{PESfit2}
\end{figure}

The functional form of the potential energy surface developed by Patridge and Schwenke is:\cite{Partridge1997:4618}
\begin{equation}
	V(r_1, r_2, \theta) = V^a(r_1) + V^a(r_2) + V^b(r_{\ff{HH}}) + V^c(r_1, r_2, \theta)
\end{equation}
where
\begin{equation}
	\begin{aligned}
	V^a(r) &= D [ e^{-2a(r-r_0)} - 2e^{-a(r-r_0)}]\\
	V^b(r) &= A e^{-br} \\
 	V^c(r_1, r_2, \theta)& = c_{000} + e^{-\beta[(r_1-r_e)^2 + (r_2-r_e)^2]} \\
 						   &\times \sum\limits_{ijk} c_{ijk}[(r_1-r_e)/r_e]^i [(r_2-r_e)/r_e]^j\\
 						   &\times[\cos(\theta) - \cos(\theta_e)]^k
 	\end{aligned}
\end{equation}
Here $r_e$ and $\theta_e$ are fixed in advanced to match water's geometry and $A$, $D, a$, $b$, $r_0$, and $c_{ijk}$ are all free parameters. As in the work of Partridge and Schwenke we truncate the polynomial expansion of $V^c$ at $i+j\leq 8$ and $k\leq 14-(i+j)$ for a total of 245 $c_{ijk}$. We found that fitting this PES to DFT monomer data was the most technically challenging part of implementing our method. The fit was performed with a training set of DFT energies for 1,176 monomer configurations. As was done by Partridge and Schwenke, we found that we had to fit to points calculated on a nonlinearly spaced grid, with more points where the PES changes rapidly (ie. around $r_{\ff{OH}} = 0.95 \Ang$). More specifically, we computed DFT energies at $r_{\ff{OH1}}, r_{\ff{OH2}} \in \lbrace$ 0.65, 0.75, 0.85, 0.95, 0.975, 1.0, 1.05, 1.1, 1.2, 1.3, 1.5, 1.6, 1.7$\rbrace \Ang$ and $\theta_{\ff{HOH}} \in \lbrace 85, 95, 100, 105, 110, 115 \rbrace$. While Partridge and Schwenke computed their fit on a grid of points going out to only $1.4\ \Ang$, we found we had to add additional points out to $1.7\ \Ang$ to obtain the correct asymptotic behaviour in the fit. Fitting to only $1.4\ \Ang$ led to occasional water dissociation events in the simulation which would cause the simulation to fail.
The results of our fitting are visualized in figures \ref{PESfit} and \ref{PESfit2}.

\subsection{Integration of the forces}
The use of the monomer PES introduces a split between intramolecular and intermolecular forces. Tuckerman et al.\ show how to derive an integration scheme when there is a splitting between long range and short range forces.\cite{Tuckerman1992:1990, tuckerman2008statistical} The method is based on the classical propagator $e^{iL \Delta t}$, which exactly evolves the system from an initial phase space point $\Gamma(t) = \lbrace \sum \bs{r}, \sum \bs{p} \rbrace$ at time $t$ to a final point at $t + \Delta t$ through $\Gamma(t + \delta t) = e^{i L \Delta t}\Gamma(t)$. The part of the Liouville operator that evolves momentum ($L_p$) is decomposed into short range ($s$) and long range ($l$) components:
\begin{equation}
	\begin{aligned}
		L &= L_\gamma + L_r + L_p^s + L_p^l  \\
		  &= L_\gamma + \sum_i^{n_a} \frac{\bs{p}_i}{m_i}\cdot \frac{\partial}{\partial \bs{r}_i} + \sum_i^{n_a} \bs{F}_i^s \cdot \frac{\partial}{\partial \bs{p}_i} + \sum_i^{n_a} \bs{F}_i^l \cdot \frac{\partial}{\partial \bs{p}_i}
	\end{aligned}
\end{equation}
Here $F_i^l$ and $F_i^s$ are the long and short range forces on atom $i$, $n_a$ is the number of atoms, $L_r$ is the part of the operator which evolves position and $L_\gamma$ refers to the part of the operator which evolves the Nos\'{e}-Hoover thermostat.
To obtain an integration method, the operator is split using the Trotter formula:
\begin{equation}\label{propsplit}
   	e^{iL \Delta t} \approx   e^{iL_\gamma \frac{\Delta t}{2}}  e^{iL_p^l \frac{\Delta t}{2}}  e^{iL_p^s \frac{\Delta t}{2} }e^{i L_r \Delta t} e^{i L_p^s \frac{\Delta t}{2}}  e^{i L_p^l \frac{\Delta t}{2}} e^{iL_\gamma \frac{\Delta t}{2}}
\end{equation}
This expression can be translated into an algorithm by reading the sequence of propagators from right to left - ie.\ $e^{iL_p^l \frac{\Delta t}{2}}$ corresponds to an half timestep update of the momentum using the long range forces, etc.\cite{Tuckerman1992:1990} The integration algorithm obtained is equivalent to a nested Velocity-Verlet scheme. Multiple time steps (MTS) can be introduced by further splitting the inner part of equation \ref{propsplit} so the short range forces are integrated $M$ times for every time the long range forces are integrated:
\begin{equation}\label{MTSprop}
   	e^{iL \Delta t} \approx e^{iL_\gamma \frac{\Delta t}{2}}  e^{iL_p^l \frac{\Delta t}{2}}
   			\left[ e^{iL_p^s \frac{\Delta t}{2 M} }e^{i L_r \frac{\Delta t}{M}} e^{i L_p^s \frac{\Delta t}{2M}}\right]^M e^{i L_p^l \frac{\Delta t}{2}} e^{iL_\gamma \frac{\Delta t}{2}}
\end{equation}
The inner timestep becomes $\frac{\Delta t}{M}$, where $M$ is an integer. Lehr et al.\ have demonstrated how MTS can be implemented in Hartree-Fock calculations for water clusters by splitting the long and short range forces via a fragment-based approach, thus showing that MTS can be done in the context of an ab initio simulation. \cite{Luehr2014:084116}  
When using MTS, one should be aware that resonances can occur between the fast timestep and the slower timestep. The first resonance occurs when the outer timestep becomes larger than $\Delta t_{\ff{max}}= \tau/\pi$, where $\tau$ is the period of the fastest mode in the problem. For water, this would be the OH stretch frequency $\approx 3600$ cm$^{-1}$ which leads to a value of $\Delta t_{\ff{max}} = 2.95 $ fs. However, in PIMD simulation one must also consider the maximum frequency normal mode of the ring polymer when combined with the maximum OH stretching frequency, which is $\sqrt{\omega_{\ff{RP,max}}^2 + \omega_{OH,max}^2}$. Our testing showed that the size of the outer timestep cannot go above $\approx$ 1.5 fs -- any longer and the simulation quickly becomes unstable. However, Morrone, et al.\ have shown that the use of colored noise thermostats can stabilize resonances, offering the possibility of higher outer timesteps.\cite{Morrone2011:014103}

\begin{figure}\centering
  \includegraphics[width=8.25cm]{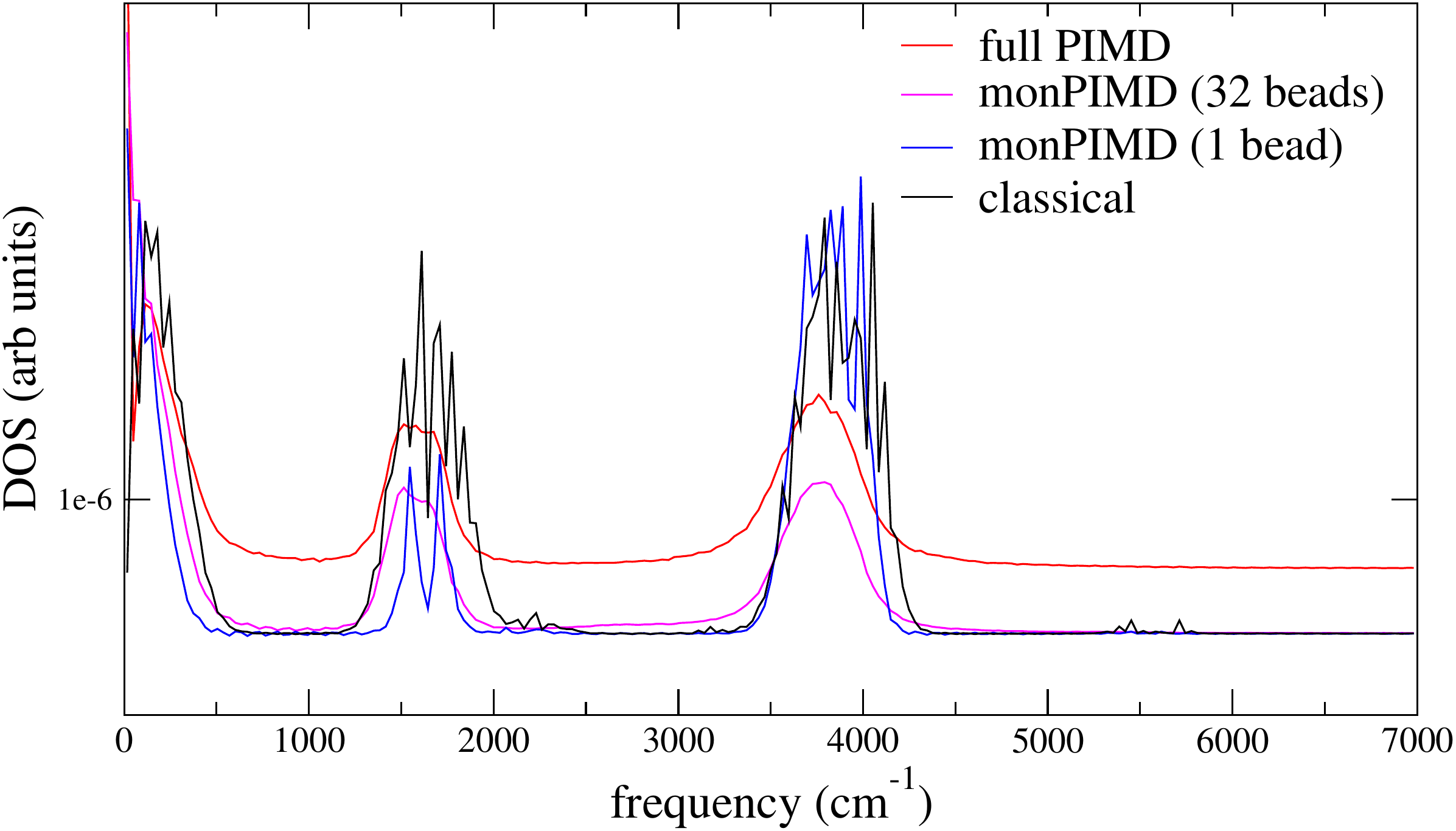}
  \caption{Validation with TTM3F: hydrogen density of states (DOS) for one molecule (gas phase) at 300 K.}
  \label{TTM3FDOS1mol}
\end{figure}

\section{Comparison to other methods}
Our method can be understood as an extension to ab initio MD of the ring polymer contraction method introduced by Markland and Manoloupolos for classical MD.\cite{Markland2008:024105,Fanourgakis2009:094102} In ring polymer contraction, long-range forces are analyzed using a contracted ring polymer with $n'$ beads that are constructed by taking the $n'$ lowest frequency ring polymer normal modes in Fourier space and transforming them into real space. Short range forces are analysed on all $n$ beads. Our method corresponds to contraction of the long range forces to $n=1$, ie.\ the centroid mode (sometimes denoted as $n=0$), and a separation between long range and short range forces that corresponds to intermolecular and intramolecular forces.

Recently, two separate groups have published a method called ``quantum ring polymer contraction'', which uses an auxiliary potential to perform ab initio PIMD with little added cost.\cite{john2015quantum,Marsalek20015:054112,Marsalek2017:1545} The method they employed, while couched in different language, is similar to the method we present here. The principal difference is that they use self consistent charge density functional tight binding (SCC-DFTB) as the auxiliary potential in place of the monomer PES we use here. As discussed before, the use of a PES makes our simulation more accurate, while using SCC-DFTB has the opposite effect.

Recently a number of papers have been published that combine ring polymer contraction with a MTS integrator and the idea of mixing forces\cite{Anglada2003:055701} from higher level and lower level ab initio methods.\cite{Marsalek20015:054112,Steele2016,Kapil:2016:054111} In such methods, a lower level ab initio technique is used to handle the short timestep and full ring polymer, while a higher level (more expensive) technique is used with the longer timestep and contracted ring polymer. For example, in two recent studies, MP2 was combined with DFT in this manner to study small gas phase molecular systems.\cite{Steele2016,Kapil:2016:054111}  A variation of this method called multilevel sampling has also been introduced and applied to FCC hydrogen, resulting in a 3-4x speedup in PIMD simulation.\cite{Geng2015:299}

Another methodology introduced recently, ring polymer interpolation, achieves a 2.5 - 10x speedup, depending on the accuracy one desires.\cite{Buxton2017:224107} Ring polymer interpolation could be combined with our method, resulting in a multiplicative speedup. Finally, another option for speeding up PIMD simulation is to incorporate adaptive resolution PIMD methods,\cite{Poma2010:adaptiveres} which allow for PIMD simulation of a small region to be combined with classical simulation of a larger region.\cite{Kreis2017:244101}

\begin{figure}\centering
  \includegraphics[width=8.25cm]{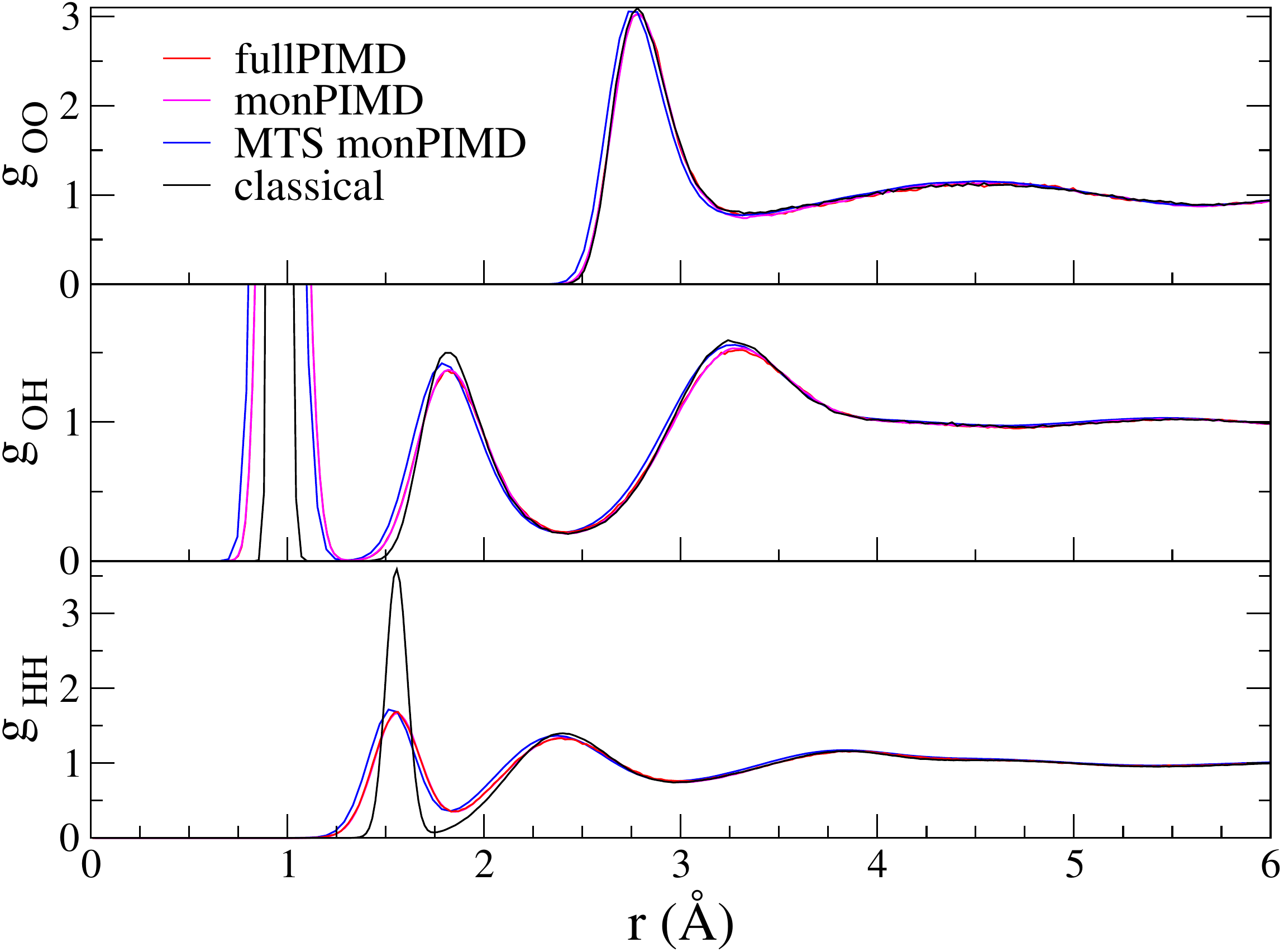}
  \caption{Validation with TTM3F: RDFs for the three methods at 300 K.}
  \label{TTM3FRDFs}
\end{figure}
\begin{table}  \centering
  \caption{Average OH distance, average HOH angle, average dipole moment, diffusion constant, average radius of gyration of the beads, max bead-bead OH distance and max centroid-centroid OH distance. Note: average OH distances for PIMD simulation are reported in the form centroid-centroid distance/bead-bead distance.}   \label{PIMDgeometry}
  \begin{tabular}{c | c c c }
   							&\multicolumn{3}{c}{TTM3F}\\
property & class. & fullPIMD & monPIMD \\
\hline
$\langle r_{\ff{OH}}\rangle$& .986    &.994/1.01   & .996/1.0  \\ 
$\langle\theta_{\ff{HOH}}\rangle$&105.43&105.4      &  105.66  \\ 
$\langle\mu\rangle$	   	    & 2.757   &  2.835      & 2.855   \\ 
$D$	(10$^{-5}$ cm$^2$/s	)	& 2.7     &  3.0        &  2.9    \\ 
$\langle r_{\ff{gyr}}\rangle$&  0.0   &  0.1507      & 0.1515          \\ 
max bead $r_{\ff{OH}}$		&  1.18  &   1.54       &  1.56           \\ 
max cent. $r_{\ff{OH}}$		&  1.13  &   1.18       &  1.23      \\ 
  \end{tabular}
\end{table}
\section{Verification of the method}

To verify that our method captures nuclear quantum effects with minimal losses in accuracy compared to a full PIMD simulation, we compare several observables - RDFs, dipole moments, density of states, the average bead radius of gyration, and OH distance histograms.
The infrared spectrum is calculated using:\cite{Ramirez2004:3973}
\begin{equation}
	n(\omega)\alpha(\omega) = \frac{\omega^2}{6 k_B T \epsilon_0 Vc} \int\limits_{-\infty}^{\infty} e^{-i\omega t} \langle \bs{P}(0) \cdot \bs{P}(t) \rangle  dt
\end{equation}
Here $\alpha(\omega)$ is the IR absorption coefficient per unit length, $n(\omega)$ is the index of refraction, and $P$ is the dipole moment of the entire system. In PIMD simulation there are two ways to calculate the dipole moment -- the first is to use the centroid positions:
\begin{equation}
    \begin{aligned}
         \bs{\mu}_i    &= \bs{\mu}(\bar{\bs{r}}_{\ff{O}}, \bar{\bs{r}}_{\ff{H1}}, \bar{\bs{r}}_{\ff{H2}} ) \\
        \bar{\bs{r}}_i &= \frac{1}{N_b} \sum_{j=1}^{N_b} \bs{r}_i^j
    \end{aligned}
\end{equation}
Here $r_{\ff{X}}$ refers to the position of atom $\ff{X}$, while $r_i^j$ refers to the position of bead $j$ in atom $i$. The second is to calculate the dipole moment separately for each bead ``image" and then average them:
\begin{equation}\label{dipolemethod2}
      \bs{\mu}_i = \frac{1}{N_b} \sum_{j=1}^{N_b} \bs{\mu}(\bs{r}_{\ff{O}}^j, \bs{r}_{\ff{H1}}^j, \bs{r}_{\ff{H2}}^j )
\end{equation}

For a linear dipole function the results are the same, but for a non-linear dipole function, such as in TTM3F or DFT, the results are not guaranteed to be the same.
In practice no difference is observed between the two methods.\cite{WittPhD} We implemented the second method (eqn.\ \ref{dipolemethod2}) because of its simplicity and because it is more in line with how estimators typically work in CMD. To calculate dipole moments for our DFT simulations, we calculated approximate dipoles using TTM3F ( a polarizable model) using the centroid coordinates from DFT.\cite{Vega2015:1145} We found that polarization included in the TTM3F dipole model is necessary to correctly capture the intensity of the OH-stretching peak.

We calculate the ``density of states'' for hydrogen using the velocity-velocity autocorrelation function:
\begin{equation}\label{DOScalc}
	I(\omega) = \frac{1}{N_H} \int\limits_{-\infty}^{\infty} e^{-i\omega t} \sum\limits_{i=1}^{N_H} \langle \bs{v}^H_i(0) \cdot \bs{v}^H_i(t) \rangle dt
\end{equation}

The extent of delocalization of the hydrogen atoms is quantified through the radius of gyration, which is the root mean square displacement of ring polymer beads from the center of the ring:
\begin{equation}\label{RofGyration}
	r_{\ff{gyr,H}} = \frac{1}{N_H N_b} \sum_{i=1}^{N_H}\sum_{j=1}^{N_b} ||\bs{r}_i^j - \bs{r}_i^c ||
\end{equation}


\begin{figure}\centering
  \includegraphics[width=8.25cm]{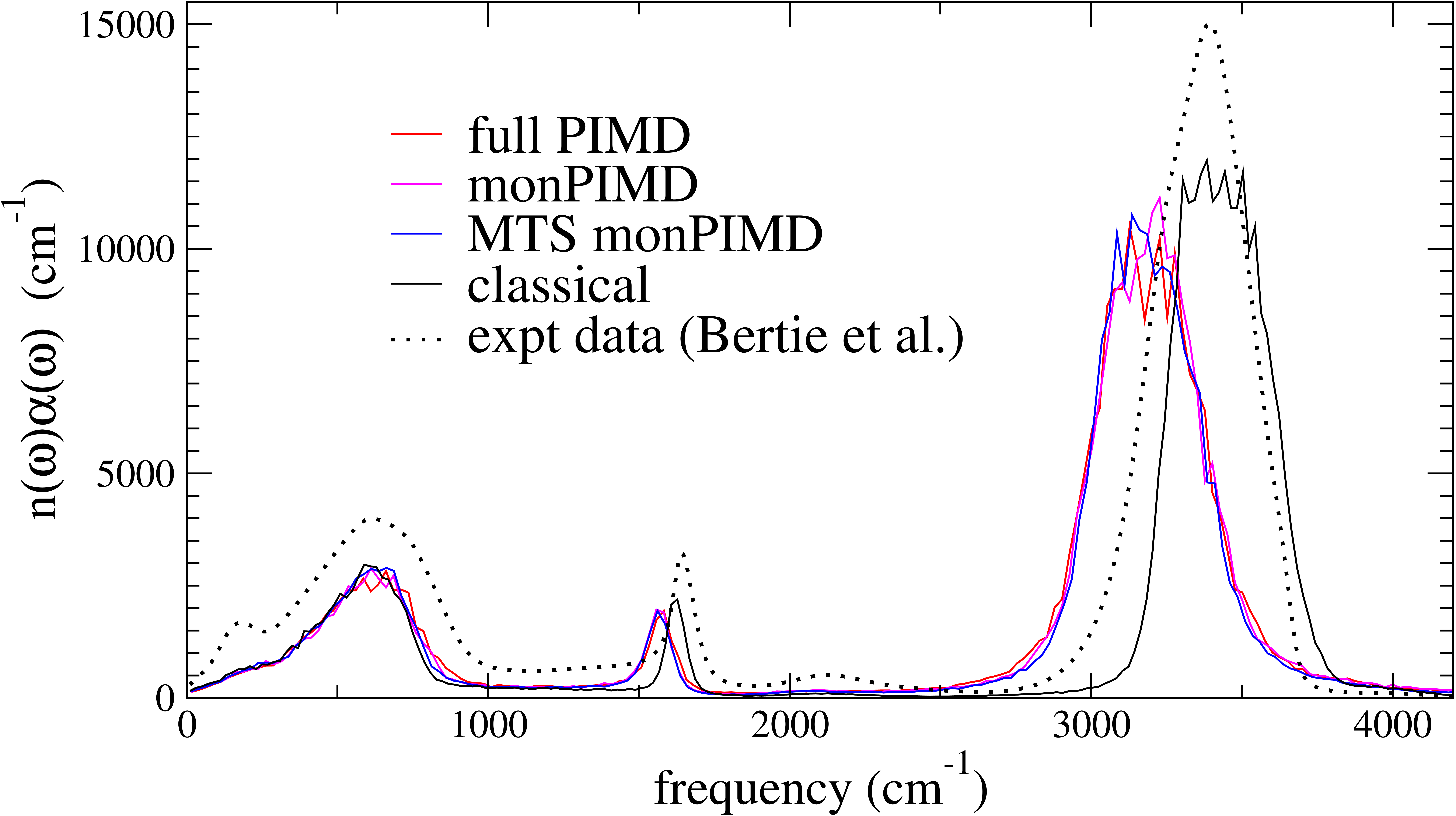}
  \caption{Validation with TTM3F: infrared spectra for the three PIMD methods for 128 molecules compared to the classical spectra and experimental data at 300 K.\cite{Bertie:96}}
  \label{TTM3Finfrared}
\end{figure}
\begin{figure}\centering
  \includegraphics[width=8.25cm]{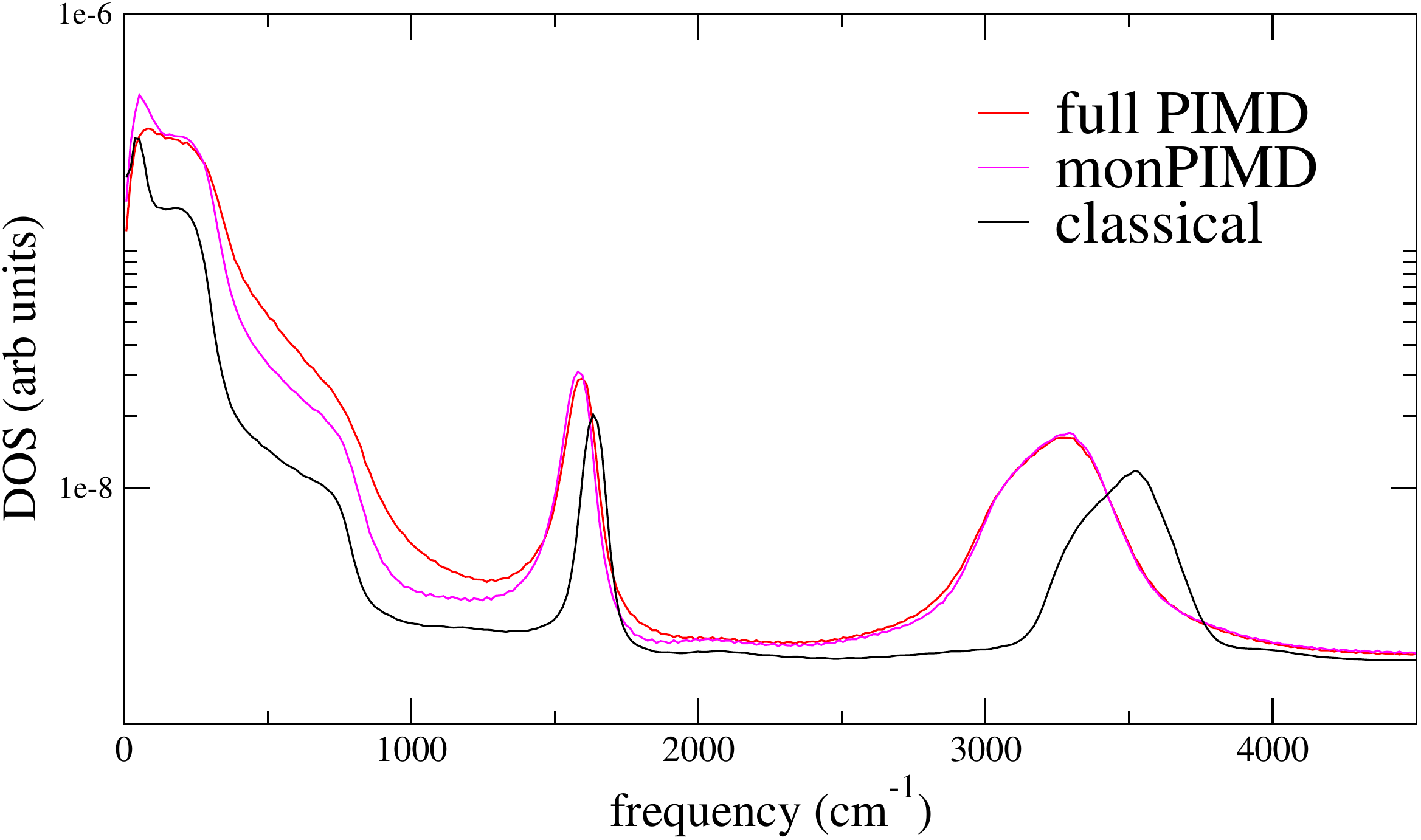}
  \caption{Validation with TTM3F: hydrogen density of states (DOS) for 128 molecules at 300 K.}
  \label{TTM3FDOS}
\end{figure}
\begin{figure}[h]
   \centering
 \includegraphics[width=8.25cm]{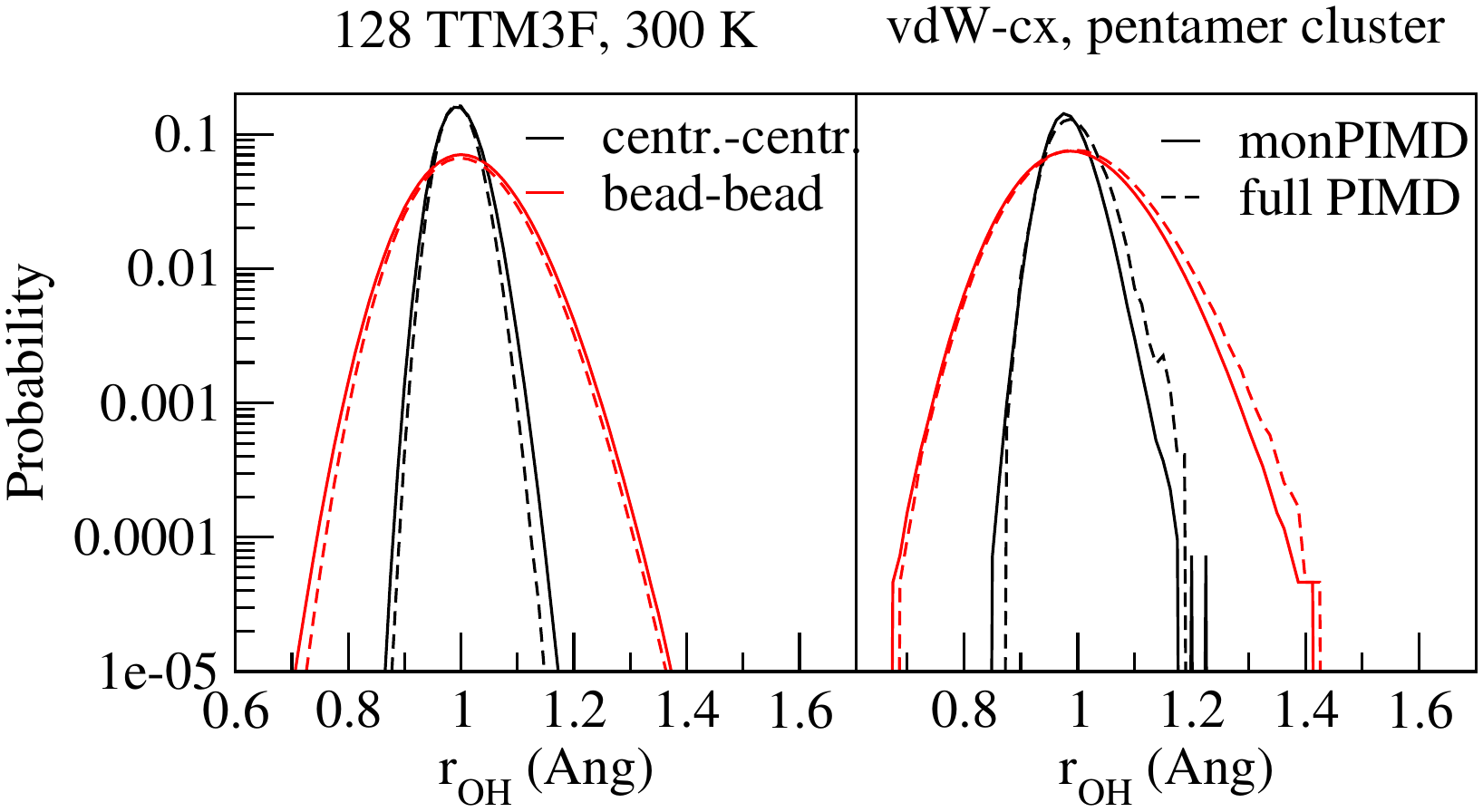}\\
  \caption{Histograms of the $r_{\ff{OH}}$ distance for a simulation of bulk water (128 molecules) with TTM3F and of a pentamer cluster with vdW-cx. Only slight differences are observed between full PIMD (solid lines) and the monomerPIMD method (dashed). }
  \label{histograms}
\end{figure}
\subsection{Tests with TTM3F}

The first verification of our method was done with the polarizable TTM3F potential, which is parametrized from ab initio simulations and uses the PS potential energy surface natively, but modified to give the correct dissociation behaviour at large $r_{\ff{OH}}$.\cite{fanourgakis:074506} We simulated a system of 256 molecules for 200 ps with a 9 $\Ang$ realspace Coulomb cutoff. Radial distribution functions (RDFs) are shown in fig.\ \ref{TTM3FRDFs}. As has been noted elsewhere, TTM3F exhibits only small primary isotope effect and very little or no secondary isotope effect,\cite{Pamuk2012:193003} due to a lack of anharmonicity in the $r_{OH}$ potential and competing quantum effects. Thus, the first O-O peak is only slightly lower and the nuclear quantum effects primarily manifest themselves in the broadening of the first O-H peak and decreased length of the second O-H peak, which indicates slightly shorter/stronger H-bonds. The monomer PIMD and full PIMD O-O RDFs are nearly the same, but the multiple time step monomer PIMD is noticeably shifted to smaller distances. The reason for this discrepancy is not clear, but very similar discrepancies are observed by Marsalek, et al.\ when applying their quantum ring polymer contraction method to RevPBE+D3.\cite{Marsalek20015:054112}

The infrared spectrum for TTM3F is shown in fig.\ \ref{TTM3Finfrared}. Since some of the parameters of TTM3F, such as the dipole moment surface, are specifically tuned to reproduce the infrared spectrum at 300 K, the placement of the peaks in the classical simulation is quite good. When NQEs are incorporated, the OH-stretching band is redshifted and broadened. The HOH bending mode is also redshifted. Our monPIMD method reproduces the PIMD spectrum almost exactly, indicating very good capturing of NQEs. Further properties are given in table \ref{PIMDgeometry}. The diffusion constant of TTM3F is only slightly increased by NQEs due to competing quantum effects, as was previously discussed for TIP4P/2005f.\cite{Habershon2009:024501} The nuclear delocalization, as measured by the radius of gyration was 1.54 $\Ang$ for the full PIMD simulation and 1.56 $\Ang$ for the monPIMD simulation. The max $r_{\ff{OH}}$ during the entire simulation, as measured by the centroid-centroid distance, was 1.18 $\Ang$ for the full PIMD simulation and 1.23 $\Ang$ for monPIMD simulation. A more complete comparison of the bead-bead delocalization in full and monomer PIMD is obtained by looking at the histograms in fig.\ \ref{histograms}. Together, the results in table \ref{PIMDgeometry} and histograms in fig.\ \ref{histograms} indicate that the delocalization in the full and approximate methods are nearly the same.

\begin{figure}\centering
 \includegraphics[width=8.5cm]{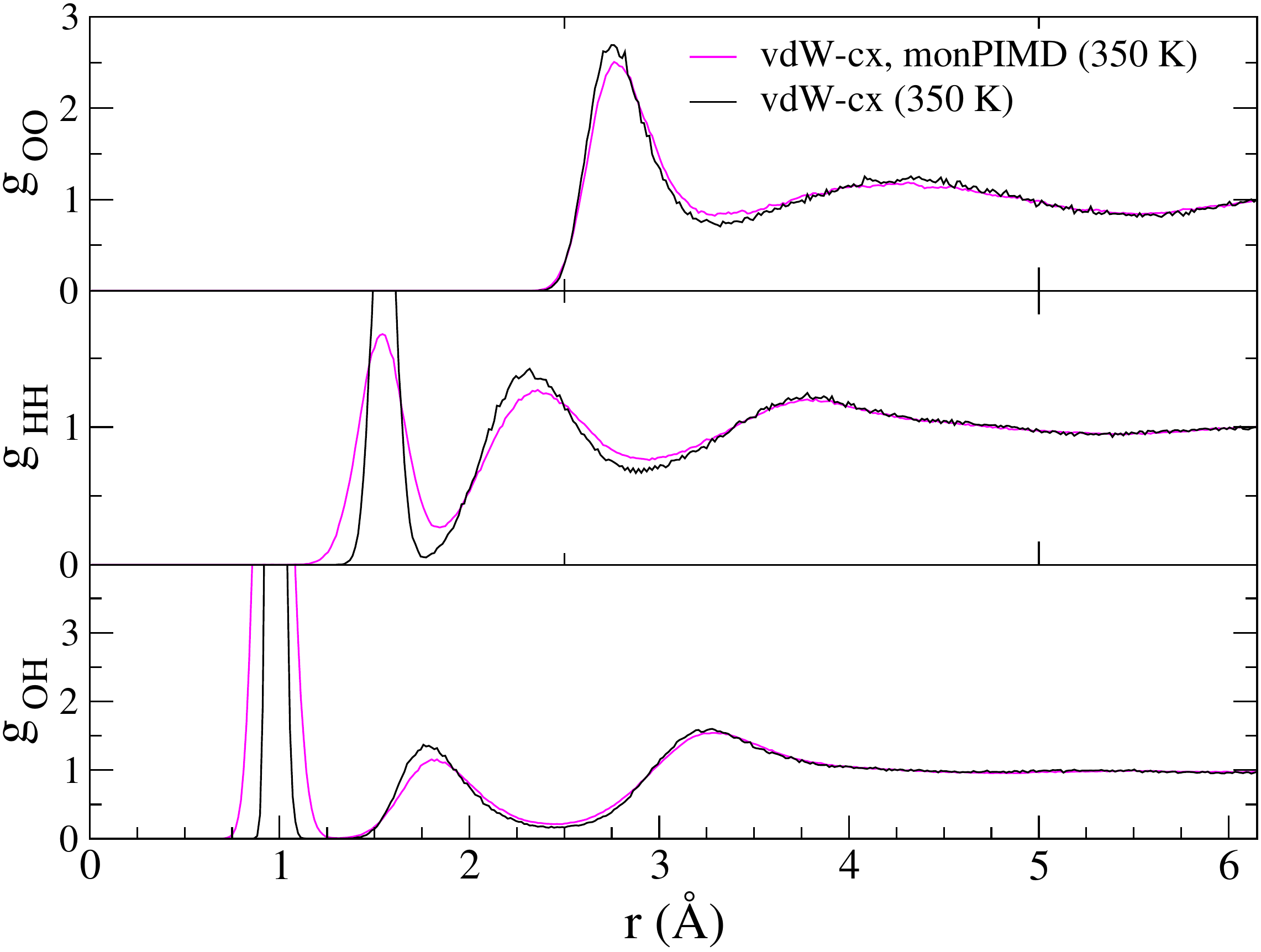}
  \caption{Comparison of RDFs for vdW-cx simulated at 350 K with the monomer PIMD method (with the monomer correction) compared to a conventional vdW-cx simulation.}
  \label{moncorrectBHcomparison}
\end{figure}

\subsection{Tests with DFT}
\begin{figure}[h]\centering
 \includegraphics[width=8.5cm]{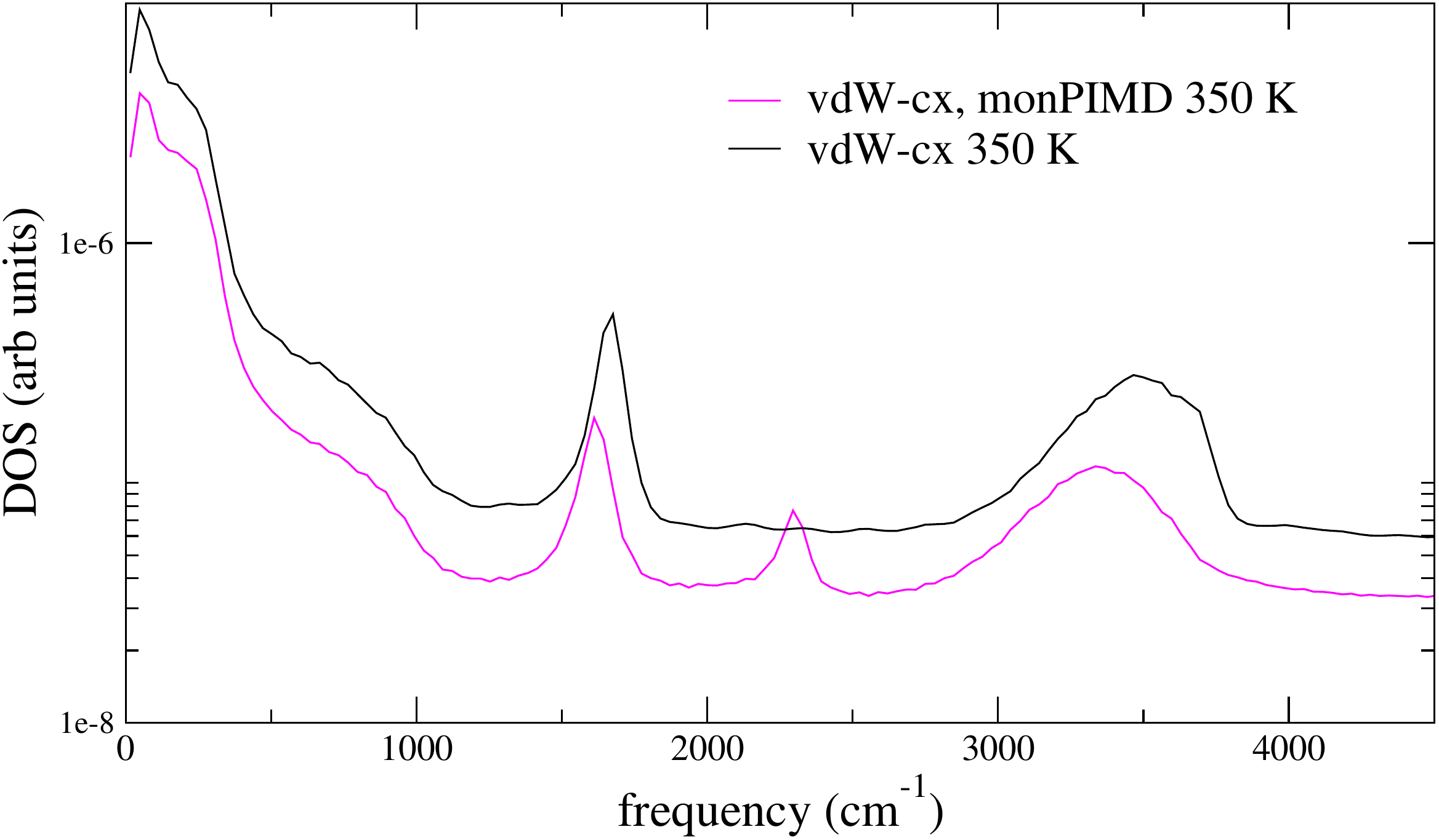}
  \caption{Density of states (eqn.\ \ref{DOScalc}) for 64 molecules with conventional MD, compared with the monomer PIMD method (32 beads). }
  \label{BH64DOS}
\end{figure}

We tested our method with PBE\cite{Perdew1996:3865} and the Berland-Hyldgaard functional,\cite{Berland20014:035412} which is a version of the DRSLL vdW functional introduced by Dion et al.\ with modified exchange.\cite{Dion2004:246401}
We choose this functional because of its consistent-exchange semilocal exchange choice (vdW-cx), which makes it very robust for simulating a variety of physical systems.\cite{Berland20014:035412} In addition, the functional performance on liquid water, has been analyzed in detail\cite{Fritz2016:224101} and shown to be comparable to other vdW-based density functionals. We began by simulating isolated molecules with both full PIMD and monPIMD, and then progressed to simulating a pentamer cluster. The distributions of $r_{\ff{OH}}$ for the pentamer cluster simulations of vdW-cx with both full PIMD and monPIMD are shown in fig.\ \ref{histograms}. The distributions of centroid-centroid and bead-bead $r_{\ff{OH}}$ distances are nearly the same, with slightly more delocalization observed in the full PIMD simulation as compared to monomer PIMD. Similar results were observed for PBE.

\begin{figure}
   \centering
 \includegraphics[width=8.5cm]{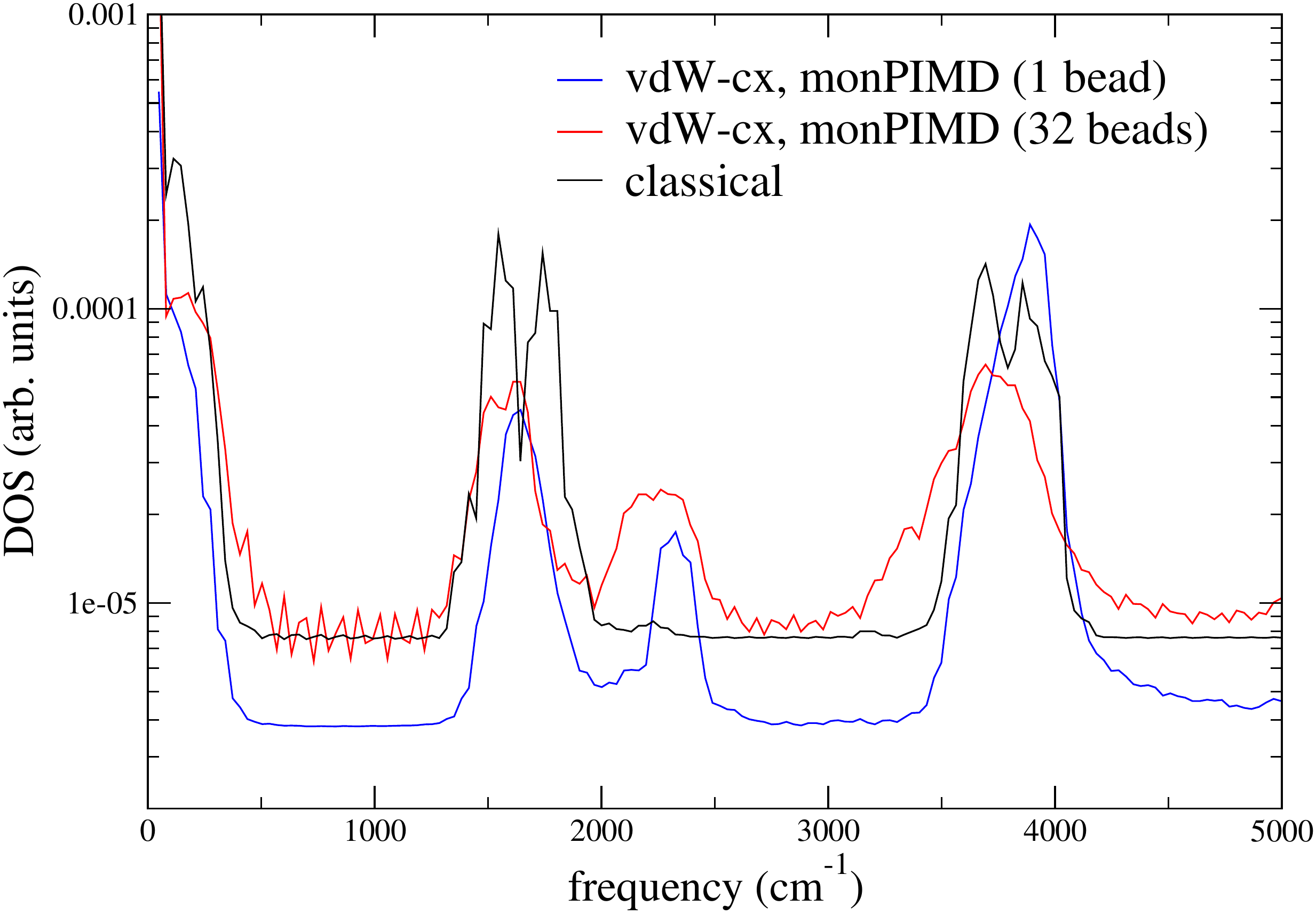}
  \caption{Density of states (eqn.\ \ref{DOScalc}) for a single molecule with the vdW-cx functional simulated with traditional classical DFT and the monomer PIMD method with 1 bead and 32 beads at 350 K.}
  \label{monPIMDmonomerBH}
\end{figure}

Figure \ref{monPIMDmonomerBH} shows the DOS for a single molecule simulated using the vdW-cx functional with conventional PIMD and our monomer PIMD method with 1 bead and 32 beads. The expected redshifting of the bending and stretching bands is observed, however additional peaks are observed at $\approx$ 2250 cm$^{-1}$ and $\approx$ 5250 cm$^{-1}$. These frequencies correspond to the association band and the first overtone band, respectively. The association band also appears in the DOS in TTM3F simulation of bulk water (see fig.\ \ref{TTM3FDOS}) but has a tiny magnitude. The spurious enhancement of both peaks is observed both with 1 beads and 32 beads, indicating it stems from some aspect of the effective potential energy surface rather than bead normal mode contamination. Careful inspection of our fit potential energy surface did not reveal any irregularities. Attempts to refit the surface with more data points were not successful in reducing the intensity of either peak in the spectra. 
Interestingly, the intensity of the association band at $\approx$ 2100 cm$^{-1}$, which is due to a combination of libration and HOH bending, has been found to be very sensitive to the coordinates used to construct the dipole moment surface and other factors such as H-bonding configuration.\cite{McCoy:2014}
Given the fact that PIMD is only rigorous for the calculation of equilibrium properties,\cite{Hone:154103} and that many methods suffer from spurious peaks from normal mode contamination,\cite{Rossi2014:234116,Habershon2008:074501} the presence of enhanced peaks in the spectrum is not as large of an issue as it may appear.

Next we performed a simulation of 64 molecules with the monomer PIMD method for both vdW-cx and PBE. A comparison of RDFs is shown in fig.\ \ref{moncorrectBHcomparison} for vdW-cx. We observe the correct destructuring of the first O-O peak and first O-O valley as well as the expected destructuring of the the O-H and H-H peaks. Information on the average water molecule geometry, dipole moment, and diffusion constant is shown in table \ref{PIMDgeometryBH}. Our simulation with monPIMD results in a slightly larger $r_{\ff{OH}}$ and HOH angle, and leads to a slightly smaller (-0.5 \%) dipole moment, and larger diffusion constant. The density of states for the 64 molecule vdW-cx simulation is shown in fig.\ \ref{BH64DOS}. Again we see that the same enhancement of the association band observed with the monomer.

\begin{table}\centering
  \caption{Comparison of properties in classical vs monPIMD simulation of 64 H$_2$O molecules with the vdW-cx functional.
  Average OH distance, average HOH angle, average dipole moment, diffusion constant, average radius of gyration of the beads, max bead-bead OH distance and max centroid-centroid OH distance. Note: average OH distances for PIMD simulation are reported in the form centroid-centroid distance/bead-bead distance. }  \label{PIMDgeometryBH}
  \begin{tabular}{c | c c  }
property     			          & class. & monPIMD           \\      
\hline
$\langle r_{\ff{OH}}\rangle$ ($\Ang$)& .994   & .986/.997         \\ 
$\langle\theta_{\ff{HOH}}\rangle$    &104.6   & 105.0/104.80    \\ 
$\langle\mu\rangle$	 (D) 	         & 3.68   &	3.66           \\ 
$D$	(10$^{-5}$ cm$^2$/s)	         &  2.3   &	3.4               \\ 
$\langle r_{\ff{gyr}}\rangle$($\Ang$)&  0.0   &    0.146         \\ 
max bead $r_{\ff{OH}}$   ($\Ang$)    &  1.19  &	 1.49          \\ 
max centroid $r_{\ff{OH}}$ ($\Ang$)  &  1.19  &	 1.19          \\ 
  \end{tabular}
\end{table}

\section{Conclusion}
We have introduced a new methodology for speeding up PIMD simulation with density functional theory and have shown that it allows for computationally tractable PIMD DFT calculations of the equilibrium properties of water. The Fortran-90 code we have written implementing this method is open source and available on GitHub.\cite{PIMDF90github} In principle our method can be applied to any molecular system. The main hurdle to applying our method to other molecules is fitting an accurate potential energy surface, however recent work has shown how this can be done with neural networks.\cite{Yao2017:014106, WangDeePMD} 

Our method was fully validated for TTM3F simulation of water, where we showed that the method reproduces both the structure and dynamics of liquid water observed in full PIMD simulation. The advantage of our method is the $\approx$ 30 x speedup obtained, which makes ab initio PIMD simulations of water practical. The method nonetheless requires careful mapping and fitting of a monomer potential energy surface whenever the DFT functional or basis set is changed. However, this process is fast and easy to implement.
While the structural properties of water found with our method are almost as good as the full PIMD simulation, we do observe enhancement of the association peak in the DOS for the ab initio functionals. We explored some possible causes for this effect and attempted to mitigate it, but more work is needed to fully understand it.

There are several variations of our method that could be explored. The first is to use a monomer DFT calculation to subtract off the monomer energies and forces (eqn.\ \ref{monomersubtr}) and then perform monomer DFT calculations to obtain forces and energies for the monomer PIMD calculations. Doing this requires $(N_b + 1) \times N_{\ff{mol}}$ additional DFT monomer calculations to be performed each timestep but has the benefit avoiding the need for a PES and providing a more accurate representation of the DFT forces and energies. We estimate there should be at least a 2x speedup over conventional PIMD with such a method, and possibly much higher. With such a method the monomer calculations can be trivially parallelized over many nodes on a cluster.

\section*{Conflicts of interest}
There are no conflicts to declare.

\section*{Acknowledgements}
We thank the Institute for Advanced Computational Science at Stony Brook University for making the LIRED and Handy clusters available to complete the simulations in this work.

\section*{Funding}
This work was partially funded by U.S. Department of Energy Award No. DE-FG02-09ER16052 (D.C.E.) and by U.S. Department of Energy Early Career Award No. de-sc0003871 (M.V.F.S.)

\bibliography{combined_references.bib}
\bibliographystyle{apsrev}

\end{document}